# Local Antiferromagnetic Correlations and $d_{x^2-y^2}$ Pairing

D.J. Scalapino

*Department of Physics*
*University of California*
*Santa Barbara, CA 93106–9530*

S.A. Trugman

*Theory Division*
*Los Alamos National Laboratory*
*Los Alamos, NM 87545*

ABSTRACT: The high $T_c$ cuprate superconductors doped near half-filling have short range antiferromagnetic correlations. Here we describe an intuitive local picture of why, if pairing occurs in the presence of short-range antiferromagnetic correlations, the orbital state will have $d_{x^2-y^2}$ symmetry.

The parent state of the high-temperature superconducting cuprates is an antiferromagnetic insulator. When holes are doped into these systems, they become metallic with short-range antiferromagnetic correlations, and below a superconducting transition temperature, the holes form singlet pairs.[1,2] There has been much discussion regarding the orbital symmetry of such pairs.[3,4,5,6] Here we consider a model with short-range repulsive interactions doped near half-filling and discuss a simple local argument which provides an intuitive picture of why, in the presence of short-range antiferromagnetic correlations, the doped holes form pairs with $d_{x^2-y^2}$ symmetry.[7,8]

For $d_{x^2-y^2}$ pairing, the momentum dependence of the gap is

$$\Delta_p = \Delta_0(\cos p_x - \cos p_y), \tag{1}$$

and the operator for adding a pair of holes has the form

$$\begin{aligned}\Delta_d =& \frac{1}{N}\sum_p \Delta_p c_{p,\uparrow} c_{-p,\downarrow} \\ =& \frac{1}{N}\sum_\ell \frac{\Delta_0}{2}\Big[\left(c_{\ell+x,\uparrow}c_{\ell,\downarrow} - c_{\ell+x,\downarrow}c_{\ell,\uparrow}\right) - \left(c_{\ell+y,\uparrow}c_{\ell,\downarrow} - c_{\ell+y,\downarrow}c_{\ell,\uparrow}\right) \\ &+ \left(c_{\ell-x,\uparrow}c_{\ell,\downarrow} - c_{\ell-x,\downarrow}c_{\ell,\uparrow}\right) - \left(c_{\ell-y,\uparrow}c_{\ell,\downarrow} - c_{\ell-y,\downarrow}c_{\ell,\uparrow}\right)\Big]. \end{aligned} \tag{2}$$

Here $c_{p,\uparrow}$ destroys a spin up electron with momentum $p$, while $c_{\ell,\uparrow}$ destroys a spin up electron on lattice site $\ell$. Note that $x$ and $y$ are unit lattice vectors so that $\left(c_{\ell+x,\uparrow}c_{\ell,\downarrow} - c_{\ell+x,\downarrow}c_{\ell,\uparrow}\right)$ creates a singlet pair between $\ell$ and an adjacent lattice site in the $x$-direction. Likewise $\left(c_{\ell+y,\uparrow}c_{\ell,\downarrow} - c_{\ell+y,\downarrow}c_{\ell,\uparrow}\right)$ creates a singlet hole pair between site $\ell$ and the adjacent site in the $y$-direction.[9] The operator given by Eq. (2) creates a superposition of these singlets around each site, giving a state with zero center of mass momentum. The key feature associated with the $d_{x^2-y^2}$ symmetry is the relative phasing $(+ - +-)$ of these singlet pairs. Here we seek to understand why holes added to a nearly half-filled band with local antiferromagnetic correlations form pairs with the relative phases given in Eq. (2). There have been various arguments as to why two holes would tend to occupy neighboring sites.

In a strong coupling approach, locating the holes on adjacent sites is favored because it reduces the number of broken exchange bonds. However, this description does not provide insight into the relative $d_{x^2-y^2}$ phasing, which we believe is an essential feature of the pairing.

Consider a system with local antiferromagnetic correlations such as the one-band Hubbard model on a square lattice. The filling is $1 - x$ electrons per site, where $x$ is small. A four-site placquette extracted from the lattice is shown in Fig. (1). The two-electron ground state on this four-site cluster, with the Hubbard interaction $U = 0$, is

$$|\psi_2\rangle = N \left( c_{1,\downarrow}^\dagger + c_{2,\downarrow}^\dagger + c_{3,\downarrow}^\dagger + c_{4,\downarrow}^\dagger \right) \left( c_{1,\uparrow}^\dagger + c_{2,\uparrow}^\dagger + c_{3,\uparrow}^\dagger + c_{4,\uparrow}^\dagger \right) |0\rangle, \qquad (3)$$

where $N$ is a normalization factor and $|0\rangle$ is the zero-particle vacuum. Equation (3) describes one up and one down electron, each with momentum $k = 0$. All of the amplitudes in Eq. (3) are positive. It is easy to verify that if a nonzero Hubbard $U$ is added to the Hamiltonian, increasing the short-range antiferromagnetic correlations, or if staggered magnetic fields are added to simulate the exchange fields of the spins surrounding the square, all amplitudes remain positive, although they no-longer have the same magnitude. The true ground state is then given by Eq. (3) multiplied by a Jastrow factor. The wavefunction, as expected, is an *s-wave* singlet. This can be seen by writing

$$|\psi_2\rangle = N \left( c_{2,\downarrow}^\dagger c_{1,\uparrow}^\dagger + c_{4,\downarrow}^\dagger c_{1,\uparrow}^\dagger + \ldots \right) |0\rangle. \qquad (4)$$

The (1,2) amplitude has the same sign as the 90° rotated (1,4) amplitude. One can thus write

$$|\psi_2\rangle = \Delta_s^\dagger |0\rangle, \qquad (5)$$

where $\Delta_s^\dagger$ is an operator that creates an s-wave pair.

The interesting point is that this same two-particle ground state $|\psi_2\rangle$ that is created by an *s*-wave operator adding particles to the vacuum, can also be created *by a $d_{x^2-y^2}$-wave* *operator* removing particles from the Mott-Hubbard insulating state with one particle per site. For the model system with four sites, $\langle\psi_2|\Delta_d|\psi_4\rangle$ is large, where $|\psi_4\rangle$ is the exact ground state with four electrons on the square. In contrast, $\langle\psi_2|\Delta_s|\psi_4\rangle = 0$.[10] It is important for this result that the state $|\psi_4\rangle$ has local antiferromagnetic correlations.[11]

It is easy to verify numerically that $\langle\psi_2|\Delta_d|\psi_4\rangle$ is large. We now motivate analytically why this is true. For a repulsive $U$ the largest real space amplitudes in the four-particle ground state wavefunction are for the "Neel" configurations

$$|\phi_a\rangle = c^\dagger_{4,\downarrow}c^\dagger_{2,\downarrow}c^\dagger_{3,\uparrow}c^\dagger_{1,\uparrow}|0\rangle, \qquad (6)$$

and the spin reversed state

$$|\phi_b\rangle = c^\dagger_{3,\downarrow}c^\dagger_{1,\downarrow}c^\dagger_{4,\uparrow}c^\dagger_{2,\uparrow}|0\rangle. \qquad (7)$$

We again examine the relative phase for an electron pair on sites (1,2) and the 90° rotated pair on sites (1,4). Annihilating the appropriate electrons, one sees that

$$c_{3,\uparrow}c_{4,\downarrow}|\phi_a\rangle = (c_{3,\uparrow}c_{4,\downarrow})\left(c^\dagger_{4,\downarrow}c^\dagger_{2,\downarrow}c^\dagger_{3,\uparrow}c^\dagger_{1,\uparrow}\right)|0\rangle = -c^\dagger_{2,\downarrow}c^\dagger_{1,\uparrow}|0\rangle, \qquad (8)$$

and

$$c_{3,\uparrow}c_{2,\downarrow}|\phi_a\rangle = (c_{3,\uparrow}c_{2,\downarrow})\left(c^\dagger_{4,\downarrow}c^\dagger_{2,\downarrow}c^\dagger_{3,\uparrow}c^\dagger_{1,\uparrow}\right)|0\rangle = +c^\dagger_{4,\downarrow}c^\dagger_{1,\uparrow}|0\rangle. \qquad (9)$$

Thus, to have a nonzero overlap against the state $|\psi_2\rangle$ of Eq. (3), one must use the operator

$$\Delta_d = (c_{3,\uparrow}c_{2,\downarrow} - c_{3,\uparrow}c_{4,\downarrow} + \ldots) \qquad (10)$$

on the four-electron Mott-Hubbard insulating state. Because of the minus sign, this is a $d_{x^2-y^2}$-wave operator.

To see that $\Delta_d$ is a *singlet* as well as a $d_{x^2-y^2}$-wave operator, we operate on the linear combination $(|\phi_a\rangle + |\phi_b\rangle)$. Note that in the four-particle ground state, $|\phi_a\rangle$ and $|\phi_b\rangle$ enter with a relative $+$ sign. This can be seen numerically, or by noting that one can reach $|\phi_b\rangle$

from $|\phi_a\rangle$ without interchanging any fermion with another of the same spin. The linear combination that gives the same relative sign as Eq. (3) is therefore obtained using

$$\left[\left(c_{3,\uparrow}c_{2,\downarrow} - c_{3,\downarrow}c_{2,\uparrow}\right) - \left(c_{3,\uparrow}c_{4,\downarrow} - c_{3,\downarrow}c_{4,\uparrow}\right) + \ldots\right]\left(|\phi_a\rangle + |\phi_b\rangle\right), \qquad (11)$$

which is the $d_{x^2-y^2}$ pair operator of Eq. (2).

In summary, to create a two particle strong-coupling bound state, when most of the "background" sites are empty, one uses an s-wave operator. However, in the experimentally relevant regime for the copper oxide superconductors, where most of the "background" already contains electrons with local antiferromagnetic correlations, a $d_{x^2-y^2}$-wave operator is required.

### Acknowledgments

We would like to thank Nejat Bulut and Jan Engelbrecht for valuable conversations. The work at Los Alamos was performed under the auspices of the U.S. Department of Energy. DJS acknowledges support from the Department of Energy under grant DE–FG03–85ER45197 and the Program on Correlated Electrons at the Center for Materials Science at Los Alamos National Laboratory.

antiferromagnetic zero-hole ground state and the doped two-hole ground state. See D. Poilblanc, *Phys. Rev. B* **48**, 3368 (1993).

**Figure Captions**

1. A local cluster of four sites, taken from a square lattice.

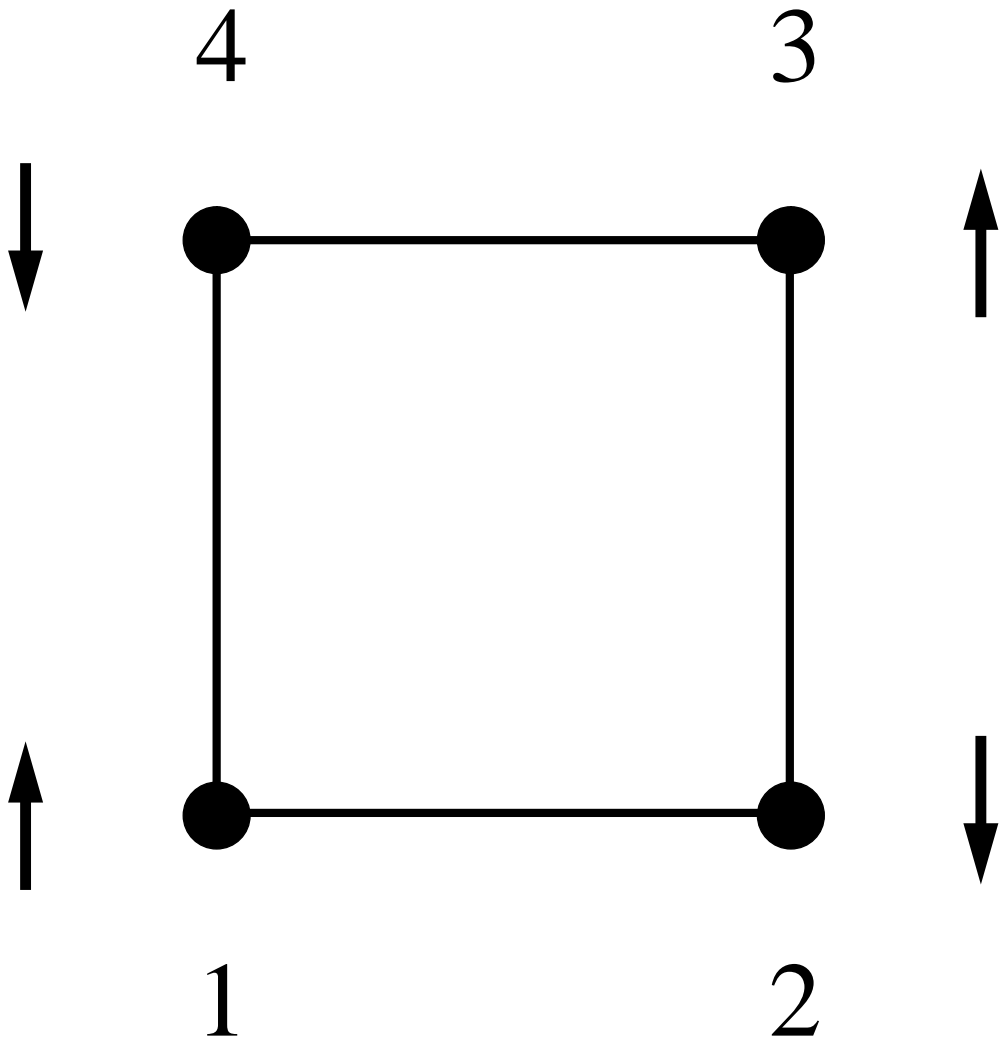